# Automatic Impact-sounding Acoustic Inspection of Concrete Structure

Jinglun Feng[1], Hua Xiao[1], Ejup Hoxha[1], Yifeng Song[2], Liang Yang[1], Jizhong Xiao[1]

[1] CCNY Robotics Lab, The City College of New York, USA, 10031
[2] Chinese Academy of Sciences, Shenyang Institute of Automation, China
email: jfeng1, ehoxha, lyang1, jxiao@ccny.cuny.edu, hxiao000@citymail.cuny.edu, songyifeng@sia.cn

ABSTRACT: Impact sounding signal has been shown containing information about the flaws of structural integrity and subsurface objects from previous research. As a non-destructive testing (NDT) method, one of the biggest challenges in impact-sounding based inspection is the subsurface targets detection and reconstruction. To address this issue, this paper presents the importance and practicability of using solenoid to trigger impact sounding signal and using acoustic data to reconstruct underground objects. First, by taking advantage of Visual Simultaneous Localization and Mapping (V-SLAM), we could obtain the 3D position of the robot during the inspection. Second, our NDE method is based on Frequency Density (FD) analysis for the Fast Fourier Transform (FFT) of the impact sounding signal. At last, by combining the 3D position data and acoustic data, this paper creates a 3D map to highlight the possible subsurface objects. The experimental results demonstrate the feasibility of the method.

KEY WORDS: Impact Sounding; Robotic Inspection; Acoustic Data; Frequency Density Analysis.

## 1 INTRODUCTION

The civil infrastructure (e.g., buildings, bridges, tunnels, dams, concrete towers) in the United States is reaching its life expectancy and the cost of inspection and repair is estimated to reach $2.9 trillion over the next 50 years [1]. Report of the Federal Highway Administration (FHWA) indicates that more than 12% of all bridges (which are 72000 bridges) in US are structurally deficient, which leads to significant public concerns and financial issue to keep these bridges in healthy condition. It is critically important to increase the inspection frequency of civil infrastructure to maintain the structural integrity of infrastructure and conduct rehabilitation operations in a timely manner. The inspection of civil infrastructure is a time-consuming, expensive, and labor-intensive task. To inspect the structural integrity of civil structures, the inspectors need to detect subsurface defects (i.e., cracks, delamination, voids) using NDT instruments such as ground penetration radar (GPR) [2], seismic pavement analyzer (PSA) [3,4], hammer sounding [5], impact sounding devices, etc., in addition to visual inspection of surface flaws.

Since most of the civil structures are made of concrete, many different NDT sensors could be used as the inspection tools. [6,7] points that GPR equipment is being used to locate many different things: from cracks in ice sheets and dams to sewage or utility pipes to metallic rebars. Ultrasound also could be used to evaluate wall of building by measuring the signal amplitude of the ultrasound through the media [8]. Impact-echo, invented by the U.S. National Bureau of Standards and Cornell University [9,10], could also be used for evaluating concrete and masonry structures [11].

However, with the current NDT inspection methods, it is still difficult to access certain inspection areas especially for subsurface area. In this paper, we focus on using impact sounding inspection method to detect subsurface area. Imapct sounding is triggered by solenoid and it is composed of striking the surface of the concrete and receiving the resulting sound. From the pattern in the impact-sounding waveform and PSD of its signal, we can indicate the existence and locations of the subsurface targets. According to the [9], the response of impact-sounding could be analyzed by using Fourier transform technique since it is dominated by P-wave sounding echoes. In the meanwhile, by analyzing the power accumulation ratio [12], the sound intensity ration [13] and the area of interval PSD [14], the PSD of acoustic signal frequency could be used as the signal features to be researched. However, impact sounding analysis is very sensitive to noise which will make it be unreliable in practical applications. In order to solve this issue, [15] implements Deep Neural Network (DNN) as the classifier for impact-acoustic signal analysis; Sarmiento [16] also represents a impact-sounding inspection method by converting the impact-sounding data into spectrum and classified it by using the inception v3 model. However, DNN method also has the drawbacks which are that it needs a large amount of training samples, depends largely on the empirical principles, and also the characteristics of the impact acoustic features suppress the generalization ability of DNN.

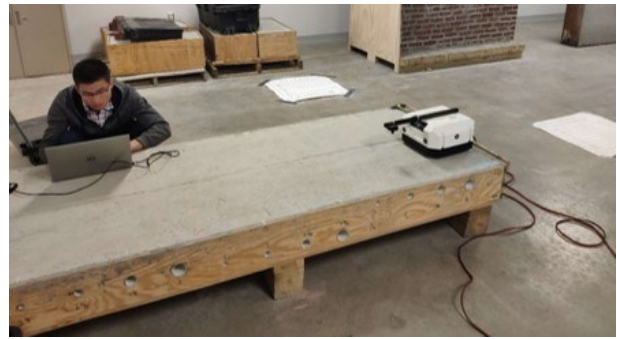

Figure 1. Picture of our inspector and the robot inspecting a slab at a test field.

The approach proposed in this paper is an automatic impact sounding robot system for the inspection of concrete slab as illustrated in Figure 1. The aims of this paper are twofold. The first aim focuses on automatic robotic inspection method in order to free the most cumbersome aspects for inspector and reduce the inspection duration. The second is the creation of a comprehensive representation of the impact-sounding results, this paper aims at creating a 3D imaging for underground objects by using robot localization results and acoustic results. In this paper, the proposed system is evaluated for concrete slab inspection. However, the proposed approach is not limited to above inspection and could be adapted to more general structures. In section II, the design of the proposed system is introduced. In section III, the theoretical basis of the impact sounding signal analysis and DNN based signal processing are introduced. In section IV, the experimental results are demonstrated, and finally the conclusion of this research is discussed.

## 2 IMPACT-SOUNDING INSPECTION SYSTEM

### 2.1 Visual pose tracking

In order to localize pipes in the scanned structure we need the pose of each data point. When we combine acoustic detection and pose, we can triangulate data points and obtain depth of the pipe. The last but not least important reason why we use SLAM is that we use information obtained to generate global acoustic inspection map. To obtain a better pose information, we used Intel D435i Realsense which has an IMU integrated.

We first initialize our system by using our previous work [17] on V-SLAM to generate visual pose. V-SLAM takes synchronized RGB image and depth image as inputs and outputs the pose of the camera; also, outputs 3D map of the environment.

There are few approaches to solve V-SLAM problem [18], we chose feature-based approach. For each RGB frame $i$ we perform feature detection $F_i = \{f(I_i^{RGB}, x_i, y_i) | i = 0,1,2, ...\}$, and using pinhole camera model and additional depth image, we have 3D information of the feature; then we perform feature detection $F_j = \{f(I_j^{RGB}, x_j, y_j) | j = 0,1,2, ...\}$ on the next RGB frame $j$. After we have the features on both images, we match corresponding features $M_{i,j} = match(F_i, F_j)$.

$$\begin{bmatrix} x_{im} \\ y_{im} \\ 1 \end{bmatrix} = Mext \begin{bmatrix} X \\ Y \\ Z \\ 1 \end{bmatrix}_{int} \tag{1}$$

Given initial pose and intrinsic parameters Mint of the camera we can estimate pose after each frame which can be achieved using (1), where $(x_{im}, y_{im})^T$ are pixel coordinates and $(X, Y, Z, 1)^T$ homogeneous coordinates of that pixel point on 3D. From all this information we form an equation of the form $A\vec{x} = \vec{b}$, then we solve this equation which combined with physical properties outputs the needed information to find rotation matrix $R$ and translation vector $\vec{t}$. This way we can estimate pose of the camera after each frame related to the previous frame, and by chain rule $T_{i,z} = T_{i,j}T_{j,k}T_{k,l...}T_{y,z}$ we can also know relationship between initial frame and current frame.

To reduce the drift we express our problem as a graph (2). To reduce the memory usage, we only save the keyframes. Keyframe, consists of pose and image frames and is introduced to represent the scenario visited. Each keyframe, it is a pivot of a local area that passed a pre-defined motion threshold. Meanwhile, we detect the overlapping between keyframes, and we form an edge connection if enough overlapping exists between any two frames. Thus, we can represent the whole scenario $\frac{vertices}{edge}$ data structure, where vertices (V) denote keyframes, edge ($E$) denotes an edge.

$$G = \{V, E\} \tag{2}$$

For any two keyframe, $i$ and $j$, the edge $E_{i,j}$ is defined with equation (3)

$$E = T_{ij} = \begin{bmatrix} R_{3x3} & t_{3x1} \\ o_{1x3} & 1 \end{bmatrix} \tag{3}$$

where $R_{3x3}$ rotation matrix and $t_{3x1}$ is translation vector, that relates vertex $V_i$ and $V_j$.

After we express our SLAM problem in a graph, we use graph optimization methods to optimize the results [19,20]. There are many methods for optimization and we use Levenberg-Marquardt(LM), which is also called damped Gauss-Newton method. This method is a robust method, and even if it starts far off the optimum it will converge fast. The update step of this method is given with:

$$\hat{x}_{k+1} = \hat{x}_k - (H - \lambda I_n)^{-1}g \tag{4}$$

where $H$ is the Hessian matrix, $I_n$ - identity matrix, $\lambda$ - weight and $g$ - gradient. As we can see this method will become as gradient descent method if $\lambda \to \infty$. Hessian matrix is calculated using $H = J^T J$, where $J$ is the Jacobian. We use the optimized pose $T_{i,j} = E_{opt\{i,j\}}$ as the correction step in our VIO system.

### 2.2 Impact-sounding measurement

In order to reveal subsurface flaws in an automatic way, the impact sounding system is designed. This system includes solenoid which is used to provide the impacting action as well as microphone which is used to receive the echo sound.

as microphone which is used to receive the echo sound. It should be noted that we provide two modes to operate impact sounding module: 1) manual mode, that is, the operator chooses the location to collect acoustic measurement through the Android controller. 2) automated acoustic inspection mode, that is, we set the system to trigger the solenoid and the microphone at 0:5HZ rate. The acoustic detection and mapping algorithm will be discussed in Section. II-C.

### 2.3 Acoustic inspection

In order to achieve automated acoustic inspection system, once an impact sounding signal is received, we have to perform the following procedures to detect the subsurface objects: 1) we need to crop the impact sounding signal from the raw audio wave; 2) we propose to use fast Fourier transform (FFT) to perform frequency analysis over the signal of interest.

### 1) Automatic signal detection

Using a microphone to record the audio signal, we have to detect the start and end of the echo signal, so we could get the information we need for analyze. To do so, we must build proper time windows for the raw audio signal. The raw audio signal is a time domain wave indicating volume magnitude. We perform a two steps operation to detect the target echo wave. Firstly, we know that our echo sounding is within 2000 Hz, and we filter the original wave with a low pass filter. Secondly, the echo sounding's maximum magnitude will over 0.999, and we detect the first time tstart of the magnitude over 0.999. Then, we select the signal of interest (SOI) as tstart - 0.01s, tstart +0.3s. Thus, we can store the SOI as a 2D array, e.g. S = (t,m), where t is the time and the m is the corresponding volume magnitude.

### 2) Frequency analysis and representation

We used frequency analysis based on previous research [21-23], to perform defect detection and area classification based on FFT. We know (see in Figure 2) that the energy of the source will be absorbed by the area have the pipe below which causes the echo sounding to have a lower energy.

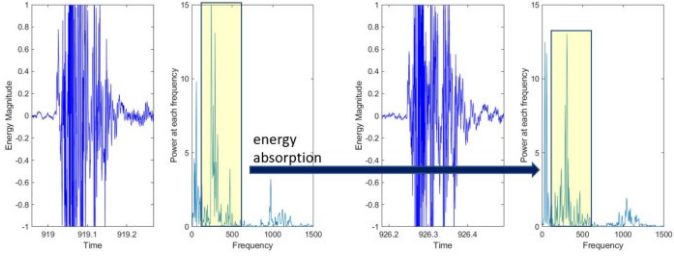

Figure 2. (a) raw impact sounding data collected on area did not have a pipe below it and it's FFT signal. (b) raw impact sounding data collected on area above the pipe, and it is FFT signal. The FFT of (a) has a much higher magnitude for frequency between 0 500HZ than that of (b), which is caused by the energy absorption of the pipe below.

Since the SOI S = (t,m) is discretized data, we directly deploy the Discrete Fourier Transform (DFT) to recover the audio using frequency and corresponding magnitude. Given the original discrete sampling data, $\{s(0), s(1), s(2), ..., \}$, we expect an approximate recover of the wave using discrete sampling [24],

$$S(f_k) = \sum_{n=0}^{N-1} x(n) \, exp(-j|2\pi f_k n) \qquad (5)$$

where $x(n) \in \{x(0), x(1), x(2), ..., x(N-1)\}$ denote the Npoint DFT magnitude, $f_k = \frac{k}{N} (k = 0,1, ..., N-1)$ denotes the sampling frequency for approximating the original wave. Then, we can obtain the FT transformation of the audio data S from time domain to frequency domain as $\{f_k, x_k\}$.

To enable visualization of the result and to quantify the acoustic measurement, in this paper we introduce frequency density (FD) representation rather than power spectral density (PSD) [25]. Given the frequency pattern $\{f_k, x_k\}$, the PSD is,

$$E_{PSD} = \sum_{i=0}^{N-1} x_i \, \delta f \qquad (6)$$

where $\delta f = f_i - f_i - 1$ However, we notice that PSD represents the area under the acoustic measurement, which means the low frequency response and high frequency response could result in the same PSD value. To solve this ambiguity, we propose frequency density (FD) to describe the energy of the acoustic measurement and the frequency serves as weight, that is,

$$E_{FD} = \sum_{i=0}^{N-1} x_i \times f_i \qquad (7)$$

## 3 ACOUSTIC INSPECTION AND 3D REGISTRATION

This paper is aiming at delivering a 3D model, with concrete defects highlighted, and the 3D model should reflect the real metric of the scenario. In this section, we discuss a machine learning approach to predict the depth and estimate the depth of the pipe under a single impact-sounding measurement. Then, we introduce a migration approach to aggregate measurements for sub-surface depth estimation.

### 3.1 Sub-surface object detection and depth prediction

An acoustic measurement is considered as a function $E: E \rightarrow R^1$, where $E$ , is the acoustic domain, and $R^1$ for a acoustic measurement that is a vector of sound magnitude. Each acoustic measurement is mapped into natural numbers first, ranging from 0 to $2^M$, where $2^M$ is the maximum. Then, we normalize to float value using the maximum value, that is, $E = \{e_0, e_1, ..., e_{n-1}\}$, where $n$ denotes the length of the array.

Once we obtain the measurement $E$, we expect we could detect whether there is a pipe buried at the current location. Besides, we also expect our algorithm can estimate the depth of the pipe. Thus, it can help us to build a 3D model to visualize the pipe and perform condition assessment.

**Intermediate Feature Extraction**

We believe both pipe detection and the depth estimation share the same encoder to extract the intermediate feature. It has been tested and proved as one of the most promising advantage for multi-modal task leaning. The intermediate feature extraction model is called Hyper Feature Model which is using the same kernel as proposed in [26]. In this paper, the Hyper Feature Model has two layers compared to a single layer represented in [26], each layer has a total 128 channels. The convolutional operation is,

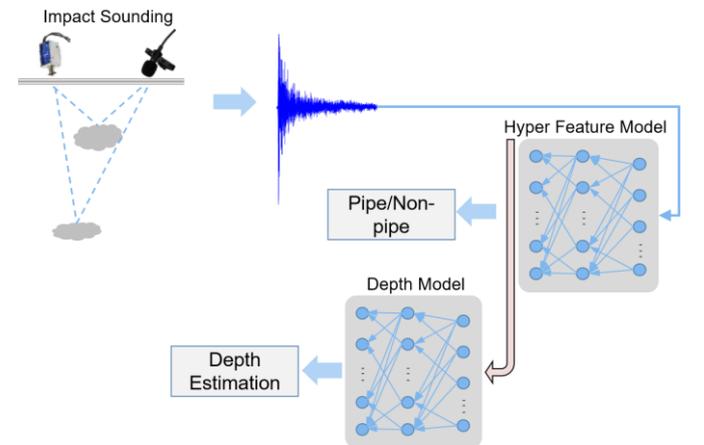

Figure 3. Model structure for depth prediction. It takes the raw acoustic signal and uses a Hyper Feature Model to extract hyperplane features. The intermediate feature is used to predict the pipe buried underground or not. Then, we develop a second model that takes the intermediate feature and estimate the depth if a pipe detected, called Depth Model.

$$Y = \sum_{i=0}^{K} (w_i e_j + b_i) \qquad (8)$$

where K denotes the total number of parameters of the kernel, $w_i$ denotes the weight of the kernel, $b_i$ is the corresponding bias, $e_j$ is an element of a segmented spectrograms. Finally, we use a maximum pooling to generate a single vector feature for pipe detection and depth estimation.

**Joint Task Leaning**

The Hyper Feature Model takes the segmented spectrograms as input, and performed 2D convolutional operation over the input to generate hyper features. To deliver our goal, the regression model is separated for pipe detection and depth estimation. In this paper, the hyper feature is represented as f(ME).

For the pipe detection, we regard this as two classes classification problem, and design a single two layered fully connected layer with 128 states and 2 states, respectively.

For the depth estimation problem, our output is only a single value, i.e., depth. Thus, we employ a $L_2$ loss, and predict the depth with float value. The Depth Model is a 3 layered model that consists of 256, 128, and 1 state respectively.

**Loss Design and Training**

It has been stated that our model takes one acoustic measurement as input and perform pipe detection and the corresponding depth estimation. Even though the prediction is separated, but we train the model in a joint approach and optimize simultaneously. The pipe detection is a two-class classification problem, and we use a cross-entropy loss as,

$$Loss(pipe) = \sum_{i=0}^{1} y_{ME,i} \, log(p(ME, i)) \qquad (9)$$

where $p(ME, i)$ is the prediction of a corresponding acoustic input.

The depth estimation is a single value estimation, and we employ $L_2$ distance loss as,

$$Loss(depth) = \|y_{ME,depth} - p_{ME,depth}\| \qquad (10)$$

where $p_{ME,depth}$ denotes the predicted depth of an acoustic input.

We finally optimize both submodels together as a joint optimization, that is, the total loss is,

$$Loss = W_0 Loss(depth) + W_1 Loss(pipe) \qquad (11)$$

It is a weighted sum of the two loss and optimize to regress the global model.

### 3.2 3D acoustic registration

In this section, this paper proposed a method that could combine the output pose information obtained from the SLAM results with acoustic data and then register them together in a 3D acoustic FD map. In Figure 4, the black lines represent the pose information provided by SLAM while all the red points represent the solenoid impact points. The color map indicates the different FD status, the brighter the color is, the more chances there is a subsurface object exists.

However, only by combining FD map with trajectory map, we still cannot predict the underground/subsurface objects. In order to get the final position prediction of the pipes which buried in the slab, we need to propose the acoustic signal to get the final results.

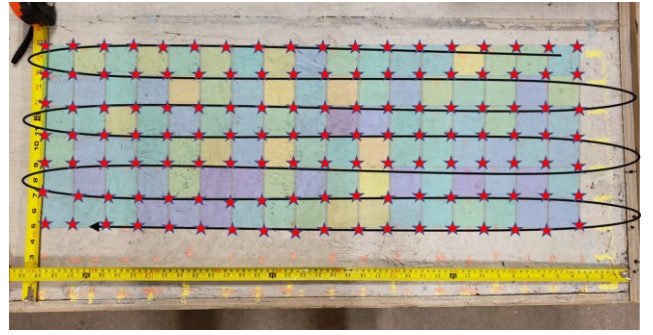

Figure 4. A 3D acoustic registration method which could combine the 3D pose information with impact-sounding acoustic information.

In this paper, to get the depth information of subsurface objects, we used Back Projection (BP) algorithm [6,27]. At each impact measurement point, Back Projection algorithm will take this point as the center and generate a semi-hemisphere with radius $r$. Radius $r$ could be calculate by extracting the peak signal in impact-sounding data, which represents the depth of the subsurface objects. Since a semi hemisphere is created, the potential target could be shown up on any points located at the surface of this semi-hemisphere. Along with the movement of impact-sounding measurement, there will be more semi-hemispheres with different radius get generated, their intersection should be the location of the targets. By this way, as shown in Figure 5, a 3D subsurface object image could be generated.

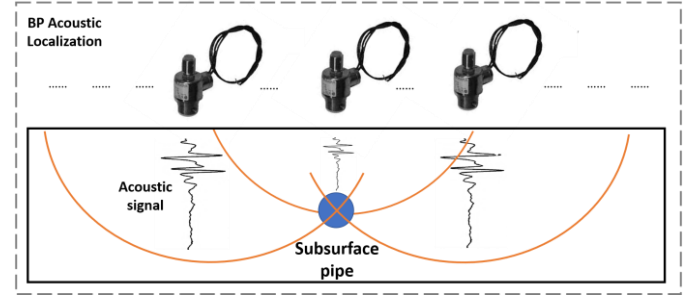

Figure 5. Back Projection algorithm implement in acoustic subsurface objects localization.

## 4 EXPERIMENTS

### 4.1 Impact-sounding data preparation

In order to perform impact-sounding inspection and target recognition, we perform field data collection at a well-designed test facility (see Figure 6 for more details). Our automated data collection system synchronized the acoustic reading $E$ and the pose estimation $P$ to obtain the coupled measurement $M$, that is, $M = (E, P)$. We collected several measurements at each location, in order to enable our training, we need to preprocess our collected measurement. For all collected data, it comes with the ground truth information including depth and length information of utility pipes and rebars.

Our data collection follows the following steps:

- Firstly, we overlay the trajectory to the 3D Testing slab model and manually label each acoustic measurement

with yes or no to decide whether there is a pipe located under the current location. Thus, we have the pairs indicated an acoustic measurement detect pipe or not as, $T^p = (x^T = E, y^T = 0)\|(x^T = E, y^T = 1)$, where use 1 to denote yes, and 0 to represent no. Meanwhile, we also annotate the distance to the nearest pipe, that is, $T^d = (x^T = E, y^T = D^{depth})$.

- Secondly, we do not learn on the raw acoustic measurement, but using segmented spectrograms as discussed in [26], that is, $ME = ms(E)$ where $ms(\cdot)$ denotes segmented mel-spectrograms operation. Each segmented measurement, we fixed the size as $60 \times 41$, and the training pairs consist of $T^p = (ME, 1|0)$, and $T^d = (ME, D^{depth})$.

- Finally, the impact sounding system is triggered by the operator to perform inspection. We collect impact sounding data at total of 126 different points, and it contains 406 sets of sounding data synchronized with pose.

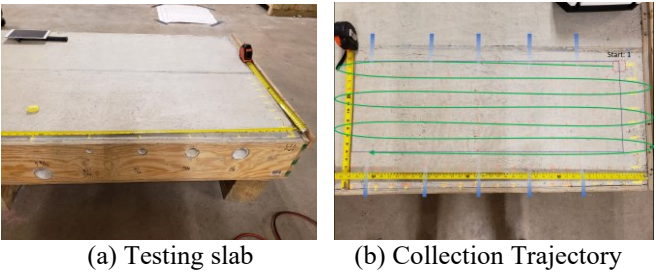

(a) Testing slab      (b) Collection Trajectory

Figure 6. Shows the testing slab area while (b) shows the robot trajectory of the data collection while the blue lines indicate the position of the buried pipes.

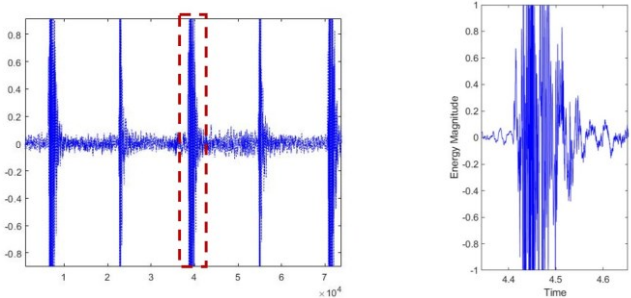

Figure 7. The left figure shows the raw data which contains 5 impact soundings at one location. The right figure shows the SoI which is cropped from raw data using method described in Section. II-C

Given a testing slab (see in Figure 6, once we collect the impact-sounding data along the trajectory shows in Figure 6 (b), we perform SOI detection as illustrated in Figure 7 Then, we perform FFT operation over the SOI. For the region classification of acoustic data, it is illustrated in Figure 8. We can clearly visualize the difference between different area via the frequency response.

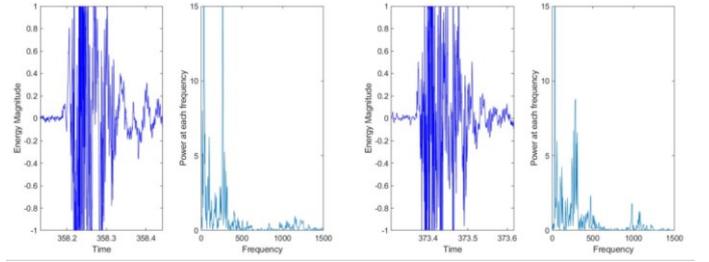

Figure 8. Region classification based on acoustic data collected at different regions, i.e., normal slab area and objects buried areas.

### 4.2 Acoustic subsurface object detection

We use Olsen solenoid as impact sounding sensor, which could be triggered by a square wave with 12V amplitude as well as a 5Hz frequency. The original audio signal should have the process of data preparation to withdraw the SOI we need. We already have the pose information using the visual pose tracking. Using this pose information, after we synchronized the acoustic reading $E$ and the pose estimation $P$, we could localize the pipes and able to create a global acoustic inspection map.

For the audio analysis, first, we divide total data into intervals of 0.75 seconds. We choose 0.75 because the intervals should be over half the time between taps. Then we set a threshold to check which intervals have relevant data. After that, we "Pads" the relevant data intervals with one interval at the end and beginning of additional time to avoid out of bounds errors. Next, we should find the index bounds of the intervals which is our SOI. Using some specific function, we designed for our signal analysis, we create envelopes for every portion of the sound data with taps detected in it. However, some envelops are just irrelevant noises. These envelops including a large amount of noise, so they were classified as "having data", we remove the areas where the maximum envelope size is over 3 standard deviations away from the mean maximum to get rid of them.

Next, we find the tap information, and add it to the data frame. We synchronized sound data, timestamps, and position data, so for every SOI area, we have their position information, time range and sound data. After figuring out the location of the tapping areas based on the real situation of our test platform, we assume the robot moved on to the next tapping location each time a large change in position (3 cm or more) is detected.

Finally, we add the pipe info as the ground truth for our analysis, so our data-frame also have the ground truth data of tap location and whether there is a pipe directly below and the distance to the nearest pipe.

Now, we have the synchronized pose and SOI data. we could use frequency density representation to visualize our result. In Figure 9, FD outputs a lower normalized energy at pipeline location, where green cylinders indicate the pipes' location.

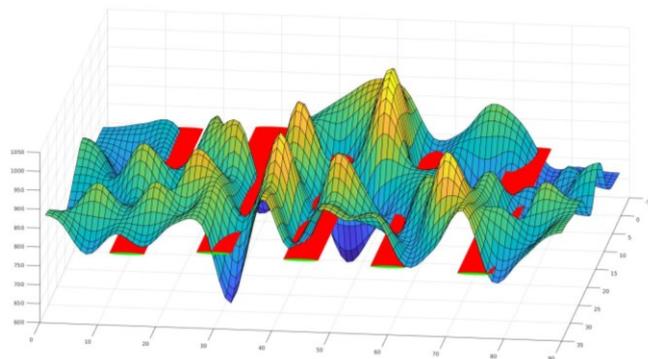

Figure 9. FD visualization result. In this figure, the color from dark to white indicates the energy difference from small to large.

## 5 CONCLUSION

This paper introduces an autonomous impact-sounding inspection system that is able to create a comprehensive representation of the subsurface inspection results. First, this system implements visual inertial fusion to estimate the pose of the solenoid. Then, based on the features extracted from the impact-sounding signal FD, an improved acoustic inspection and 3D registration method was implemented to perform the classification and target re-localization. Finally, the proposed DNN based method is used to predict the depth of subsurface objects, according to the estimation of FD distribution of the acoustic signal. The experiments show the effectiveness of our proposed 3D subsurface objects reconstruction methodology.

ACKNOWLEDGMENT

Financial support for this study was provided by NSF grant IIP-1915721, and by the U.S. Department of Transportation, Office of the Assistant Secretary for Research and Technology (USDOT/OST-R) under Grant No. 69A3551747126 through INSPIRE University Transportation Center (http://inspireutc. mst.edu) at Missouri University of Science and Technology. The views, opinions, findings and conclusions reflected in this publication are solely those of the authors and do not represent the official policy or position of the USDOT/OSTR, or any State or other entity. J Xiao has significant financial interest in InnovBot LLC, a company involved in R&D and commercialization of the technology.